\documentclass[12pt]{iopart}

\usepackage{iopams}
\usepackage{graphicx}

\newcommand{\bra}[1]{\left\langle#1\right|}
\newcommand{\ket}[1]{\left|#1\right\rangle}

\begin{document}

\title[Correlations between spectra with different symmetry]{Correlations between spectra with different symmetry: \\
 any chance to be observed?}
\author{P. Braun$^{1,2}$, F. Leyvraz$^3$ and T.H. Seligman$^{3,4}$}
 \address{ $^{1}$Fachbereich Physik, Universit\"at Duisburg-Essen,
47048
 Duisburg, Germany}
\address{ $^2$Institute of Physics, Saint-Petersburg University,
198504 Saint-Petersburg,
  Russia}
\address{$^3$Instituto de Ciencias F\'isicas, Universidad Nacional
Aut\'onoma de M\'exico,
Cuernavaca, Morelos, M\'exico}
\address{$^4$Centro Internacional de Ciencias, Cuernavaca, Morelos,
M\'exico}

\begin{abstract}
A standard assumption in quantum chaology is the
absence of correlation
between spectra pertaining to different symmetries. Doubts were
raised about this statement for several reasons, in particular,
because in semiclassics spectra of different symmetry are
expressed in terms of the same set of periodic orbits. We
reexamine this question and find absence of correlation in the
universal regime. In the case of continuous symmetry the problem
is reduced to parametric correlation, and we expect correlations
to be present up to a certain time which is essentially classical
but larger than the ballistic time.

\end{abstract}

\maketitle

\section{\protect\bigskip Introduction}

The topic we are to consider is the cross-correlation between
spectra of different symmetry for so-called quantum chaotic
systems, i.e. quantum systems, whose classical analogue is
chaotic. The instinctive reaction would be to say that states of
different symmetry live in different spaces, don't talk to each
other and cannot conceivably display any correlation
\cite{Haake10}. Doubts appear when one remembers that in the
semiclassical approach the level correlations of different
symmetries can be expressed in terms of sums over the same set of
periodic orbits, and such "brotherhood in the parent orbits" may
introduce a correlation. Although independence is habitually
assumed, the question deserves a closer look. In nuclear physics
some evidence has been assembled that correlations in two-body
random ensembles with finite sets of states exist, which is not
too surprising considering the small number of independent
two-body matrix elements \cite{Papen06}, and certain nuclear
scattering data support the existence of correlations to some
extent \cite{Flore05} but a study of correlations between spectra
of opposite parity in the limit of many particles revealed no
correlations \cite{Papen08}.

These findings make it all the more desirable to analyze the usual
assumption of independence in the semi-classical limit in some
detail, as we shall do here. We get basically a negative result,
both for the case of discrete and the continuous symmetry, i.e., the
absence of correlations is confirmed in the universal regime.

These conclusions are to be taken with a grain of
salt. Correlations may still exist in
certain pathological systems (mostly those with disconnected phase
space often
 due to unusual behavior under time-reversal symmetry \cite{Veble07, Leyvr96, Gutki07}.
Our reasoning also does not necessarily apply to arithmetic
billiards which have chaotic dynamics, but an exponentially large
number of geodesics with identical actions \cite{Bogom97}).
Nevertheless, it has been shown by explicit evaluation of the
Selberg trace sum formula for the correlation function, that
eigenvalues for even and odd parity in the modular domain on the
surface with constant negative curvature are indeed uncorrelated
\cite{Bogom96} More importantly, for continuous symmetries
correlations may exist up to a certain time which is essentially
classical in nature, if the difference of quantum numbers
characterizing the irreducible representations is of order one as
 the quantum numbers themselves get large. Note that this
effect can be easily overlooked, as times are often given in terms
of the Heisenberg time which, in the semi-classical limit, goes to
infinity.

The paper is structured as follows: After laying foundations we
discuss the discrete
 symmetries in what we consider to be the simplest
case. This is the reflection symmetry in a two-dimensional system
when we can  pass to a half-space with Dirichlet or Neumann
boundary conditions. We obtain the relevant results for these two
particular cases using periodic orbit expansions and find the
expected absence of correlations. Starting with the diagonal
approximation we discuss later the off-diagonal contributions and
times beyond the Heisenberg time. After that we study general
discrete symmetries. Trivial generalization of the described
technique is possible only for groups that induce a decomposition
of space into fundamental domains. This is not always true and we
shall present the general argument using the symmetry
decomposition of periodic orbits, as proposed in \cite{Selig94}.

In the next section we pass to continuous symmetry and consider
the effect of a change of the index of irreducible representation
as parametric correlation \cite{Faas93} within the
symmetry-reduced Hilbert space. Again, we find absence of
correlations in the universal regime, i.e., on the scale of the
Heisenberg time, however correlations on a classical i.e.
$\hbar$-independent time scale are expected. Finally we present
some conclusions and an outlook about possible observations.

\section{Definitions: spectral cross-correlation
function and the form factor}

We start with some definitions that will be essential to the later
arguments, and at the same time we fix the notations we shall use.

Let $\rho ^{(1)}(E)=\sum_{k}\delta \left( E-E_{k}^{(1)}\right) $
and $\rho ^{(2)}(E)=\sum_{k}\delta \left( E-E_{k}^{(2)}\right) $
be two spectral densities describing either two subspectra of
different symmetry of the same Hamiltonian or the spectra of two
different Hamiltonians. We shall assume that both spectra are
unfolded to the same constant mean level density $\Delta $ (see,
e.g., \cite{Haake10}).  The precise manner of the unfolding
procedure is a somewhat thorny issue and will be discussed later.

The two-level cross-correlation function is defined as,%
\begin{equation}
R^{\left( 1,2\right) }(E,\varepsilon )=\frac{\left\langle \rho
^{\left( 1\right) }\left( E+\frac{\varepsilon \Delta}{2\pi
}\right) \rho ^{(2)}\left( E-\frac{\varepsilon \Delta }{2\pi
}\right) \right\rangle } {\left\langle \rho ^{\left( 1\right)
}\left( E\right) \right\rangle \left\langle \rho ^{(2)}\left(
E\right) \right\rangle }-1 \label{correlator}
\end{equation}%
where $\left\langle \ldots \right\rangle $ denotes averaging over
an interval of $E$ and smoothing over a window of the
dimensionless energy offset $\varepsilon $; since the spectra are
unfolded we have $\left\langle \rho ^{\left( 1\right) }\left(
E\right) \right\rangle =\left\langle \rho ^{\left( 2\right)
}\left( E\right) \right\rangle =\Delta ^{-1}$.

It is often more convenient to work with the Fourier transform of
$R^{\left( 1,2\right) }(E,\varepsilon )$ with respect to
$\varepsilon$. The
result is the cross  form factor which is a double sum,%
\begin{equation}
K^{\left( 1,2\right) }\left( \tau \right) =\sum_{k,l}e^{i\left(
E_{k}^{\left( 1\right) }-E_{l}^{\left( 2\right) }\right) \tau
T_{H}/\hbar }\,; \label{defformf}
\end{equation}%
the dimensionless $\tau =T/T_{H}$ is time in units of the
Heisenberg time $T_{H}=2\pi \hbar /\Delta $.

A semiclassical representation for the spectral correlator and the
form factor follows from the Gutzwiller expansion of the
fluctuating part of the spectral density in terms of the classical
periodic orbits $\gamma $,
\begin{equation}
\rho _{\mathrm{osc}}(E)\sim \sum_{\gamma ^{\left( 1\right) }}A_{\gamma
}e^{iS_{\gamma }/\hbar -i\mu _{\gamma }\pi /2}.  \label{Gutzwiller}
\end{equation}%
Here $S_{\gamma },A_{\gamma },\mu _{\gamma }$ are the action,
stability coefficient and the Maslov index of the orbit $\gamma $.
Strictly speaking this expression should also include the periodic
orbit repetitions; these however become irrelevant in the
semiclassical limit. Replacing in (\ref{defformf}) both spectral
densities  by the respective expansions (\ref{Gutzwiller}) we get,
\begin{equation}\fl
K^{\left( 1,2\right) }\left( \tau \right) \sim \frac{1}{\delta
}\left\langle%
\sum_{
 T_{H}\tau <T_{\gamma _{1}},\,T_{\gamma _{2}}<T_{H}\tau +\delta
 }%
A_{\gamma _{1}}A_{\gamma
_{2}}e^{i\left( S_{\gamma _{1}}^{\left( 1\right) }-S_{\gamma
_{2}}^{\left(
2\right) }\right) /\hbar -i\left( \mu _{\gamma _{1}}^{\left( 1\right)
}-\mu
_{\gamma _{2}}^{\left( 2\right) }\right) \pi /2}\right\rangle ;
\label{semigen}
\end{equation}%
the brackets $\left\langle \ldots \right\rangle $ denote averaging
over $E$ and smoothing over a small window of $\tau $. The sum is
taken over periodic orbits $\gamma _{1}$ of the system $1$ and
$\gamma _{2}$ of the system $2$. Their periods $T_{\gamma
_{1}},T_{\gamma _{2}}$ must lie in the interval $\left[ T,T+\delta
\right] $ whose width $\delta $ is small compared with $T=\tau
T_{H}$.

\section{\protect\bigskip\ Discrete symmetry}
To illustrate the central point of our argument we shall start
with the simplest possible example of a system with a single
reflection symmetry.

\subsection{ A simple example: reflection symmetry}
\begin{figure}
\begin{center}
  \includegraphics[scale=.6]{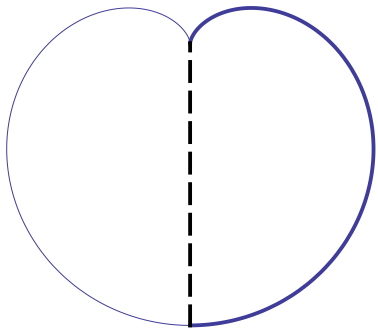}
\end{center}
\caption{Cardioid billiard. Fundamental domain is bordered by  the
 symmetry line (dashed)  and the right (bold)
half of the billiard wall.} \label{cardioid}
\end{figure}

Consider the spectral correlation between the even and odd spectra
in a chaotic system with two degrees of freedom such as the
cardioid billiard \cite{Robni83,Baeck95,Baeck97}, see
Fig.~\ref{cardioid}.
 %.
On the line of symmetry the even and odd wave functions either
have zero normal derivative or themselves turn into zero. The
subspectra of definite parity can be found separately by solving
the Schr\"{o}dinger equation in the desymmetrized billiard (the
fundamental domain) obtained by cutting the billiard along the
symmetry line and dropping one of the halves. Imposing the Neumann
(resp. Dirichlet) boundary condition on the cut we shall obtain
the even\ (resp. odd) part of the spectrum.

The semiclassical level density consists of the smooth part (the
Weyl term) and the fluctuating part which is the Gutzwiller sum
over periodic orbits of the desymmetrized billiard with the
mirror-reflecting wall at the symmetry line. The difference
between the even and odd cases in the Weyl term  is responsible
for the fact that the $N$-th eigenvalue in the symmetric sector
$E_{N}^{S}$ lies lower than the $N$-th eigenvalue in the
antisymmetric sector $E_{N}^{A}$ by an amount of the order $\sqrt
E_{N}^{A, S}$, see \cite{Pavlo94} and reference therein. This
issue has some consequences but can be taken care by an
appropriate unfolding procedure. In the Gutzwiller sum the even
and odd cases differ by the effective Maslov indices of the
periodic orbits since each reflection in the Dirichlet and Neumann
boundary contributes $\pi $ or zero, respectively, to the phase of
the orbit contribution.

We shall denote by $n_{\gamma }$ the number of strikes of the
periodic orbit $\gamma $ of the desymmetrized billiard against the
line of symmetry. It is easy to see that orbits with even
$n_{\gamma }$ correspond to the periodic orbits of the full
billiard with the same period; the latter are non-symmetric unless
$\gamma $ is self-retracing.
%, see Fig. 1.
On the other hand orbits with odd $n_{\gamma }$ are associated
with symmetric orbits of the full billiard whose period is doubled
compared with $\gamma $.
% (Fig. 2).
It is physically reasonable to
assume (although we don't know whether the result is proven) that
the number of orbits $\gamma $ with periods in a certain interval
$\left[ T,T+\delta \right] $ with $T$ large may not depend on
parity of $n_{\gamma }$, as well as on parity of the number of
visits to any side of the desymmetrized billiard.

The semiclassical cross form factor now becomes a double sum over
the same set of periodic orbits of the desymmetrized system. In
the following expression summation over $\gamma _{1}$ and $\gamma
_{2}$ is connected with the Gutzwiller expansion of the even and
odd spectral densities, respectively,
\begin{equation}
K_{gu}\left( \tau \right) \sim \frac{1}{\delta }\left\langle%
\sum_{
 T_{H}\tau <T_{\gamma _{1}},\,T_{\gamma _{2}}<T_{H}\tau +\delta
 }%
 A_{\gamma _{2}}\left(
-1\right) ^{n_{\gamma _{2}}}e^{i\left( S_{\gamma _{1}}-S_{\gamma
_{2}}\right) /\hbar }\right\rangle , \label{semiclformf}
\end{equation}%
the factor $\left( -1\right) ^{n_{\gamma _{2}}}$ emerges because
each visit to the symmetry line with the Dirichlet boundary
condition yields the Maslov phase $\pi $.

\subsection{Diagonal approximation}

Dropping in (\ref{semiclformf}) all summands with $\gamma _{1}$ not
coinciding with $\gamma _{2}$ (up to time reversal since we have
time reversal invariance)
 we come to the diagonal approximation,%
\begin{equation*}
K_{gu,\mathrm{diag}}\left( \tau \right) \sim \frac{1}{\delta
}\left\langle \sum_{{ T_{H}\tau <T_{\gamma }<T_{H}\tau
+\delta }}%
 A_{\gamma }^{2}\left( -1\right) ^{n_{\gamma
}}\right\rangle .
\end{equation*}%
It differs from Berry's diagonal approximation for autocorrelation
form factor \cite{Berry85} by the presence of the sign factor
$(-1)^{\gamma }$ leading to the destructive interference among the
contributions. Assuming that for the long orbits with very large
$n_{\gamma }$ the number of orbits with even and odd $n_{\gamma }$
is the same and not correlated with the stability coefficients we
obtain $K_{gu,\mathrm{diag}}\left( \tau \right) =0.$

A warning has to be made here. Our basic argument (even and odd
spectral densities are created by the same set of orbits which
contribute with the same action but a different Maslov index)
implicitly assumes ergodicity of the classical motion and loses
its validity if ergodicity is violated. Consider, e.g., a particle
moving in a symmetric double-well potential in two dimensions. For
energies below the dividing barrier the classical motion is
clearly non-ergodic since the available phase space disconnects
into two symmetric domains; even and odd levels in the quantum
problem are then almost degenerate, up to the tunnelling
splitting. Consequently the cross $g\longleftrightarrow u$ form
factor will practically coincide with the autocorrelation form
factor of the spectrum in a single well.

 Another trivial issue should also be pointed out here: for very
short times (ballistic times) corresponding to the period of the
shortest classical
orbits, there are no more cancellations between different orbits
and correlations certainly arise.

\subsection{Off-diagonal contributions}

The off-diagonal contribution for the auto form factor differing
from (\ref{semiclformf}) by the absence of the sign factor
$(-1)^{n_{\gamma }}$ is known to stem from pairs of the so called
orbit-partners. A partner $\gamma ^{\prime }$ of an orbit $\gamma
$ is the orbit which practically coincides with $\gamma $
everywhere but in the \textquotedblleft
encounters\textquotedblright , i.e.,\ a set of two or more orbit
stretches almost parallel and abnormally close to each other. The
lengths of $\gamma $ and $\gamma ^{\prime }$ can be so close that
their action difference can be comparable to or smaller than
$\hbar $. Consequently the corresponding contributions to the auto
form factor will avoid the destructive interference which, in the
case of randomly composed pairs, would lead to annihilation of the
contribution. The leading off-diagonal correction is due to the so
called Sieber-Richter pairs \cite{Siebe01}; higher order terms
giving the auto form factor as a series in powers of $\tau $ are
calculated in \cite{Muell05}. Analytically continuing the result
one gets the auto form factor for times smaller than the
Heisenberg time, i.e., for $\tau <1$; for larger times a different
approach is needed (see the following subsection).

Let us now address the cross form factor. Again, contributions of
randomly composed pairs are expected to average to zero.  Limiting
summation only to pairs of the orbit-partners we may reduce
(\ref{semiclformf})
to%
\begin{equation*}
K_{gu}\left( \tau \right) \sim \left\langle \frac{1}{\delta
}\sum_{{  \tau T_{H}<\overline{T}_{\gamma }<\tau
T_{H}+\delta }}\left( -1\right) ^{n_{\gamma }}%
A_{\gamma
}^{2}\left( \sum_{\gamma ^{\prime }(\gamma )}e^{\frac{i\Delta
\bar{S}_{\gamma \gamma ^{\prime }}}{\hbar }}\right) \right\rangle
.
\end{equation*}%
The inner sum is taken over all partners $\gamma ^{\prime }$ of
the orbit $\gamma $ including the diagonal ones. Since the partner
orbit coincides with the original orbit everywhere except the
relatively short encounter stretches, the Maslov indices of
$\gamma $ and $\gamma ^{\prime }$, coincide, apart from the
contribution of reflections from the Dirichlet boundary. Again, we
assume that there is no correlation between parity of the number
of strikes of the orbit $\gamma $ against the cut and the value of
the sum over partners of $\gamma $ multiplied by its stability
factor. Hence the expected value of $K_{gu}\left( \tau \right) $,
with the off-diagonal contributions taken into account, is zero,
at least for $\tau <1$.

\subsection{Times larger than the Heisenberg time}

\bigskip The fact that maxima (and in the case of GSE, even
singularities) of form factors can occur at the Heisenberg time,
makes it necessary to investigate the long-time behaviour in the
discrete symmetry case. Semiclassical investigation of spectral
auto and cross correlation beyond the Heisenberg time was started
by Bogomolny and Keating\cite{Bogom96a}. Complete expansion of the
autocorrelation functions using the formalism of the
4-determinantal generating functions is given in
\cite{Heusl07,Muell09}; the results of the latter can be
generalized  to cross-correlations.

The appropriate generating function of the two-point
cross-correlation functions for the spectra of Hamiltonians
$H^{\left( 1\right) },H^{\left( 2\right) }~$\ is a ratio of four
spectral determinants,
\begin{equation}\fl
Z^{\left( 12\right) }\left( \varepsilon _{A},\varepsilon
_{B},\varepsilon
_{C},\varepsilon _{D}\right) =\left\langle \frac{\det \left(
E+\varepsilon
_{C}\Delta /2\pi -H^{\left( 1\right) }\right) \det \left(
E+\varepsilon
_{D}\Delta /2\pi -H^{\left( 2\right) }\right) }{\det \left(
E+\varepsilon
_{A}\Delta /2\pi -H^{\left( 1\right) }\right) \det \left(
E+\varepsilon
_{B}\Delta /2\pi -H^{\left( 2\right) }\right) }\right\rangle _{E}
\label{defz}
\end{equation}%
where $\left\langle \ldots \right\rangle _{E}$ as
usual indicates averaging over the center energy $E$
and smoothing over small windows of the energy offsets
from $E$; unlike the generating function in the
problem of autocorrelation it is not symmetric with
respect to the exchange of $\varepsilon _{C}$ and
$\varepsilon _{D}$.  Using the identity
\begin{equation} \frac{\partial \ln \det \left(
E-H\right) }{\partial E}=\mathrm{Tr}\left( E-H\right)
^{-1}  \label{sdettogreen}
 \end{equation}
and
proportionality between $\mathrm{Im}\mathrm{Tr}\left(
E-H\right) ^{-1}$ and the level density $\rho \left(
E\right) $ we obtain the cross-correlation function as
the second derivative,%
\begin{eqnarray}\fl
R^{\left( 12\right) }\left( \varepsilon \right)
&=&\mathrm{Re}\lim_{\eta \rightarrow +0}\left\langle
\mathrm{Tr}\left( E+\frac{\varepsilon ^{+}\Delta } {2\pi
}-H^{\left( 1\right) }\right) ^{-1}\mathrm{Tr}\left(
E-\frac{\varepsilon ^{+}\Delta }{2\pi }-H^{\left( 2\right)
}\right) ^{-1}\right\rangle _{E}-1 \nonumber
\label{ztocorrf} \\
&=&2\mathrm{Re}\lim_{\eta \rightarrow +0}\left. \frac{\partial
^{2}}{\partial \varepsilon _{A}\partial \varepsilon _{B}}Z^{\left(
12\right) }\left( \varepsilon _{A},\varepsilon _{B},\varepsilon
_{C},\varepsilon _{D}\right)
\right\vert _{\parallel },  \\
\varepsilon ^{+ } &=&\varepsilon + i\eta\, .
\end{eqnarray}%
Here $\eta $ is a small positive number, and the symbol
\textquotedblleft
$\parallel $\textquotedblright\ denotes the substitution of the
arguments,
\begin{eqnarray*}
\varepsilon _{A} &=&\varepsilon ^{+},\quad \varepsilon
_{B}=-\varepsilon
^{+}, \\
\varepsilon _{C} &=&\varepsilon ,\quad \varepsilon _{D}=-\varepsilon .
\end{eqnarray*}%
\qquad \qquad \qquad \qquad

The semiclassical approximation for the generating function
follows from that for the spectral determinants. Using the so
called "Riemann-Siegel lookalike" asymptotics of $\det \left(
E-H\right) $ introduced by Berry and Keating \cite{Berry90,
Keati92, Berry92} one can show that the semiclassical estimate of
$Z^{\left( 12\right) }$ consists of two terms. One of these
substituted into (\ref{sumthet}) leads to a correlation function
whose Fourier transformation reproduces the form factor for small
times. The second term in $Z^{\left( 12\right) }$ is responsible
for the behavior of the form factor for times larger than $T_{H}$.
Each term in the semiclassical generating function is obtained as
an infinite product over the periodic orbits, averaged over the
central energy $E$. The leading order contribution is obtained by
the diagonal approximation which, in the formalism of the
generating functions, means that the factors associated with
different periodic orbits are assumed independent (with the
obvious exception of the mutually time-reversed orbits);
consequently averages of products over the periodic orbits are
replaced by the products of averages.

\bigskip The diagonal approximation for the generating function
describing the $gu-$correlation  is deduced  in the same way, step
by step, as in the case of autocorrelation; for the lack of space
we address the readers to the papers \cite{Heusl07,Muell09}. At a
certain stage one gets,

\begin{eqnarray*}
Z^{\left( gu\right) }\left( \varepsilon _{A},\varepsilon
_{B},\varepsilon
_{C},\varepsilon _{D}\right)  &=&Z_{I}+Z_{II}, \\
Z_{I} &=&e^{\frac{i}{2}\left( \varepsilon _{A}-\varepsilon
_{B}-\varepsilon _{C}+\varepsilon _{D}\right) }\prod_{\gamma }\exp
g \vartheta
_{I,\gamma } \\
Z_{II} &=&e^{\frac{i}{2}\left( \varepsilon _{A}-\varepsilon
_{B}-\varepsilon _{D}+\varepsilon _{C}\right) }\prod_{\gamma }\exp
g \vartheta _{II,\gamma }\,.
\end{eqnarray*}%
The factor $g$ in the exponents is equal to $2$  for systems with
time reversal allowed (like the cardioid billiard) where it
reflects the existence of pairs of mutually time reversed orbits;
in the absence of time reversal $g=1$. The exponents are given by
\begin{eqnarray*}
\vartheta _{I,\gamma }\, &=&\left\vert F_{\gamma }\right\vert
^{2}\left(
e^{i\tau _{\gamma }\varepsilon _{A}}-e^{i\tau _{\gamma }\varepsilon
_{C}}\right) \left( e^{-i\tau _{\gamma }\varepsilon _{B}}-e^{-i\tau
_{\gamma
}\varepsilon _{D}}\right) \left( -1\right) ^{n_{\gamma }}, \\
\vartheta _{II,\gamma } &=&-\left\vert F_{\gamma }\right\vert
^{2}\left[
e^{i\tau _{\gamma }\varepsilon _{A}}-\left( -1\right) ^{n_{\gamma
}}e^{i\tau
_{\gamma }\varepsilon _{D}}\right] \left[ e^{-i\tau _{\gamma
}\varepsilon
_{C}}-\left( -1\right) ^{n_{\gamma }}e^{-i\tau _{\gamma }\varepsilon
_{B}}
\right] .
\end{eqnarray*}
  We denote here by $\tau _{\gamma }=T_{\gamma }/T_{H}$ the
period of a periodic orbit $\gamma $ in terms of the Heisenberg
time; $F_{\gamma }$ is its stability coefficient. As before
$n_{\gamma }$ is the number of visits of the orbit $\gamma $ to
the cut in the desymmetrized billiard; if the factors $\left(
-1\right) ^{n_{\gamma }}$ were dropped we would return to the case
of autocorrelation.

\bigskip It remains to perform summation over the periodic orbits
in the exponents of $Z_{I},Z_{II}$. Again, we assume absence of
correlation between parity of $n_{\gamma }$, on the one hand, and
the orbit period and the stability coefficient, on the other hand.
Then the summands containing the factor $\left( -1\right)
^{n_{\gamma }}$ disappear after summation over $\gamma $ such
that%

\begin{eqnarray}
\sum_{\gamma }\vartheta _{I,\gamma } &\rightarrow &0,  \\
\sum_{\gamma }\vartheta _{II,\gamma } &\rightarrow &-\sum_{\gamma
}\left\vert F_{\gamma }\right\vert ^{2}\left[ e^{i\tau _{\gamma
}\left(
\varepsilon _{A}-\varepsilon _{C}\right) }+e^{i\tau _{\gamma }\left(
\varepsilon _{D}-\epsilon _{B}\right) }\right] ,  \label{sumthet}
\end{eqnarray}%
consequently the term $Z_{I}$ in the generating function is an
irrelevant
constant.

The sum in the second line of (\ref{sumthet}) can be  evaluated
using the sum rule of Hannay and Ozorio di Almeida\cite{Hanna84}
for the stability coefficients $F_{\gamma }$,%
\begin{equation*}
\sum_{T<T_{\gamma }<T+\Delta T}\left\vert F_{\gamma }\right\vert
^{2}= \frac{\Delta T}{T}
\end{equation*}%
which gives
\begin{eqnarray*}
\sum_{\gamma }\vartheta _{II,\gamma } &\approx &-\int_{\tau
_{0}}^{\infty }
\frac{d\tau }{\tau }\left[ e^{i\tau \left( \varepsilon
_{A}-\varepsilon
_{C}\right) }+e^{i\tau \left( \varepsilon _{D}-\epsilon _{B}\right)
}\right]
\\
&\approx &\ln \left[ \left( \varepsilon _{A}-\varepsilon
_{C}\right) \left( \varepsilon _{D}-\epsilon _{B}\right) \right]
+2\ln \tau_0+2\gamma-i\pi
\end{eqnarray*}%
with $\gamma$ denoting Euler constant. The lower integration limit
is $\tau _{0}=T_{0}/T_{H}$ with $T_{0}$ standing for some minimal
period above which periodic orbits behave approximately
ergodically; in the semiclassical limit $\tau _{0}$ tends to zero.
 The second term in the generating
function
is thus,%
\begin{equation*}
Z_{II}\left( \varepsilon _{A},\varepsilon _{B},\varepsilon
_{C},\varepsilon _{D}\right) \propto e^{\frac{i}{2}\left(
\varepsilon _{A}-\varepsilon _{B}-\varepsilon _{D}+\varepsilon
_{C}\right) }\left[\tau_0^{2}\left( \varepsilon _{A}-\varepsilon
_{C}\right) \left( \varepsilon _{D}-\epsilon _{B}\right)
\right]^g.
\end{equation*}%
The respective two-point correlation function is zero in the
presence of time reversal when $g=2$. Without time reversal
($g=1$) the term $Z_{II}$ contributes like $\mathrm{const}\,\,
\tau_0^2 e^{i2\varepsilon}$ to the cross-correlation function  or,
equivalently, like a delta-peak to the time dependent cross form
factor $K_{gu}\left( \tau \right)$ at $\tau=1$, i.e., at the
Heisenberg time. However, since the term is proportional to
$\tau_0^2$ this contribution vanishes in the semiclassical limit.
Therefore, as could be expected, correlations between the even and
odd part of the spectrum do not exist for times either smaller or
larger than the Heisenberg time.

\subsection{General discrete symmetries}
In this section we give a formal treatment of the problem in the
presence of more complicated discrete symmetry groups $G$. The
level density $\rho \left( E\right) $ falls into a sum of
subdensities $\rho _{m}\left( E\right) $ connected with different
irreducible representations (IR) $m$ of $G$. The semiclassical
approximation for the fluctuating part of $\rho _{m}\left(
E\right) $ is provided by the symmetry-adapted Gutzwiller trace
formula \cite {Selig94}. In the following, we shall follow the
ideas expounded in \cite{Selig94}, so that our treatment includes
the case in which the reduction of the configuration space $M$ of
the system to a fundamental domain $\bar{M}$, which tesselates the
full $M$ when the symmetry operations are applied to it, is
difficult or impossible.

Let us look at the cross form factor defined as in (\ref{defformf})
\begin{equation}
    K^{(m_1,m_2)}(\tau)=\sum_{k,l}e^{i\left(E_{k,m_{1}}-
    E_{l,m_{2}}\right)t/\hbar}\,;
    \label{eq:1}
\end{equation}
throughout this section the argument of the form factor
$\tau=t/T_H$, i. e., time measured in units of the Heisenberg
time. What we want to show is that this quantity vanishes. At the
abstract level this can be rewritten as
\begin{equation}
    K^{(m_{1},m_{2})}(\tau)=\left[
    \Tr\left( P_{m_1}e^{-iHt/\hbar}P_{m_1}\right)\right]^\star\cdot
    \left[
    \Tr \left(P_{m_2}e^{-iHt/\hbar}P_{m_2}\right)
    \right]
    \label{eq:2}
\end{equation}
where $P_{m}$ is the projector of the Hilbert space on the IR
$m$. It is given by the expression
\begin{equation}
    P_{m}=\frac{1}{|G|}\sum_{g\in G}\chi_{m}(g)g
    \label{eq:3}
\end{equation}
where $\chi_{m}(g)$ is the character corresponding to the IR
$m$ and $g$ denotes the group element, which is assumed to act
directly on the elements of the Hilbert space.

As is well known, traces can be estimated semiclassically
as integrals over coherent states  $\ket{p,q}$. Assuming that
the symmetry operations $g$ have a well defined semiclassical limit,
we know how they act on the phase space as well as on
the Hilbert space, and thus are able to define
such quantities as $g(p,q)$.
One therefore obtains
\begin{eqnarray}\fl
    K^{(m_1,m_2)}(\tau) & = &
    \frac{1}{|G|^2}\sum_{g,g^\prime\in G}
    \int\frac{dp_1\,dq_1\,dp_{2}\,dq_{2}}{(2\pi)^{2d}}
\left[\chi_{m_1}(g)
   \bra{p_{1},q_{1}}e^{-iHt/\hbar}\ket{g(p_{1},q_{1})}\right]^*
    \times
    \nonumber  \\
     & \times &  \chi_{m_2}(g^\prime) \bra{p_{2},q_{2}}e^{-iHt/\hbar}\ket{g^\prime(p_{2},q_{2})}
    \label{eq:4}
\end{eqnarray}
where we have used the identity $\Tr (PAP)=\Tr (AP)$ where $P$ is
an arbitrary projector. Since the $\ket{p,q}$ are well-localized
in phase space, the kind of expressions occurring in (\ref{eq:4})
only differs from zero when the phase space point $(p,q)$ is (at
least approximately) carried over to $g(p,q)$ in time $t$ by the
classical evolution. In any case, this reasoning is legitimate
when $t$ is less than the Ehrenfest time. Since, in this section,
we shall limit ourselves to the diagonal approximation, which is
only valid when times are not too large, this is a reasonable
approach. Now we must account correctly for all the various
contributions to this integral.

To this end, let us start by looking at the quantity $\Tr
\left(P_{m_1}e^{-iHt/\hbar}P_{m_1}\right)$. Let us define as a
generalized periodic orbit of period $t$ any segment which
connects a phase space point $(p,q)$ with one of its symmetric
partners $g(p,q)$ over time $t$. We call such orbits $g$-orbits,
making explicit reference to the corresponding group element. Let
$\lambda$ be a $g$-orbit. One then additionally defines
$g_{\lambda}$ to be the minimal element of the symmetry group $G$
such that $g_{\lambda}(p,q)$ lies on the orbit $\lambda$ going
through $(p,q)$ and such that no other element $g^\prime$ of $G$
has the property that $g^\prime(p,q)$ lies on the orbit $\lambda$
between $(p,q)$ and $g_{\lambda}(p,q)$. Note that in general
$g_{\lambda}\neq g$. There is, however, as shown in
\cite{Selig94}, always an integer $n$ such that $g=g_{\lambda}^n$.

Let $(p,q)$ lie on a $g$-orbit $\lambda$. One then has
\begin{equation}
    \bra{p,q}e^{-iHt/\hbar}\ket{g(p,q)}=A_{\lambda}e^{iS_{\lambda}/\hbar}
    \label{eq:5}
\end{equation}
where $A_{\lambda}$ is related to the stability of the orbit in the
usual way and the phase $e^{iS_{\lambda}/\hbar}$ contains all the
information on the action of the $g$-orbit as well as its Maslov
phases.

Integrating the l.h.s. of (\ref{eq:5}) over $p,q$ leads to
\begin{equation}
    \Tr\left(
    P_{m_2}e^{-iHt/\hbar}P_{m_2}
    \right)=\frac{1}{|G|}\sum_{g\in G}
    \frac{t}{(2\pi)^d}\sum_{n=1}^\infty\sum_{\lambda}
    A_{\lambda}e^{iS_{\lambda}/\hbar}\chi_{m_2}(g)
    \delta_{g,g_{\lambda}^n}
    \label{eq:6}
\end{equation}
where $n$ is the number of iterations of the primitive element
leading to $g$. The factor $t$ arises from the fact that an
integration over the whole segment from $(p,q)$ to $g(p,q)$ is
possible and yields the same result as starting with $(p,q)$. The
sum still runs independently over the various symmetric ``copies''
of the $g$-orbit $\lambda$, in particular over its iterations
which lead eventually to a bona fide periodic orbit. Neglecting
repetitions, that is $n\neq1$, yields
\begin{equation}
    \Tr\left(
    P_{m_2}e^{-iHt/\hbar}P_{m_2}\right)=
    \frac{t}{(2\pi)^d|G|}\sum_{\lambda}
    A_{\lambda}e^{iS_{\lambda}/\hbar}\chi_{m_2}\left(
    g_{\lambda}\right)\,.
    \label{eq:7}
\end{equation}
Let us now multiply this with the complex conjugate of the corresponding
expression for $m_{1}$. This gives:
\begin{equation}
    K^{(m_1,m_2)}(\tau)=\frac{t^2}{(2\pi)^{2d}|G|^2}\sum_{\lambda,\mu}
    A_{\lambda}A^\star_{\mu}e^{i(S_{\lambda}-S_{\mu})/\hbar}
    \chi^\star_{m_1}\left(g_{\mu}\right)\chi_{m_2}\left(g_\lambda\right)\,.
    \label{eq:8}
\end{equation}
One now introduces the diagonal approximation, which involves
keeping in this sum only such terms as have a phase equal to one, that is,
only to sum over these pairs of $\lambda$ and $\mu$ such that
$S_{\lambda}=S_{\mu}$. Since we consider the different iterations of a
$g$ orbit to be different, and since we have already discarded orbits
which are non-primitive, and since, in the same vein, we discard
such orbits as have a non-trivial symmetry (that is, we assume that
the only $h\in G$ such that $h(p,q)=(p,q)$ is the identity), it is
clear that the total number of such pairs is $|G|^2$ independently of
$\lambda$ and $\mu$: there are $|G|$ degenerate $\lambda$'s and the
same number of degenerate $\mu$'s. The final result is hence
\begin{equation}
     K^{(m_1,m_2)}(\tau)=\frac{t^2}{(2\pi)^{2d}}\sum_{\lambda}
    |A_{\lambda}|^2
    \chi^\star_{m_1}\left(g_{\lambda}\right)\chi_{m_2}\left(g_{\lambda}\right)
    \label{eq:9}
\end{equation}
where summation is carried out now over one representative $g$-orbit
$\lambda$ among the $|G|$ degenerate ones.
In other terms, the sum corresponds to a desymmetrized
problem. Keeping all $g$-orbits would require a compensating factor of
$|G|^{-1}$ in front of the r.h.s. of (\ref{eq:9}).

One now rewrites this as
\begin{equation}
     K^{(m_1,m_2)}(\tau)=\frac{t^2}{(2\pi)^{2d}}\sum_{g}
     \chi^\star_{m_1}(g)\chi_{m_2}(g)
     \sum_{\lambda}
    |A_{\lambda}|^2
    \delta_{g,g_{\lambda}}
    \label{eq:10}
\end{equation}
and we now make a final assumption altogether similar to those
that we have been making throughout in this paper: namely, we
assume that when the $g$-orbits are sufficiently long, they are
equidistributed on all elements of the group $G$. This leads to
stating that the sum over $\lambda$ is in fact independent of $g$.
This is, of course, sufficient to guarantee that
$K^{(m_1,m_2)}(\tau)=0$ whenever $m_1\neq m_2$.

\section{Continuous symmetry}

In the presence of continuous symmetry, the problem has one or
several integrals of motion due to Noether's theorem. We focus on
one of them which can be chosen as an action and whose eigenvalues
in the semiclassical limit are given by $\hbar m$, where $m$ is integer.
Our question is whether correlation may exist between the energy
spectra of two operators, $H_{m}$ and $H_{m+1}$. Here we denote by
$H_{m}$ the Hamiltonian $H$ {\em reduced\/} to the appropriate
symmetry subspace. In the following, we limit ourselves to the case in
which the dynamics of these Hamiltonians are all
overwhelmingly chaotic in the energy range of
interest.

A simple example which we
shall use to illustrate our argument is a system with axial
symmetry in a 3d space. Introducing cylindric coordinates we can
separate the angular part of the eigenfunctions writing $\Psi
=e^{im\phi }\psi \left( \rho ,z\right) $ where $\psi \left( \rho
,z\right) $ is an eigenfunction of the Hamiltonian $H_{m}\left(
\rho ,z\right) $ in the subspace of states with the given
$z$-component of the angular momentum $L_{z}=\hbar m$.
As stated above, we assume
that the $2d$ classical motion described by this Hamiltonian
 is chaotic. A moment's reflection will show that this example
 contains all the relevant features of the more general situation
 described at the beginning of this section.

In this case, the Hamiltonians $H_{m}$ and $H_{m+1}$ correspond to
the radial Hamiltonians with magnetic quantum number $m$ and $m+1$
respectively. The magnetic quantum number enters these
Hamiltonians as a parameter, and the well-known theory of
parametric correlations can be applied. Special treatment is
necessary when $m=0$ or $m\sim 1$. The difference between the two
Hamiltonians due to the centrifugal term $W=L_z^2/2M\rho^2$ ($M$
is the mass of the particle)  is then of the formal order
$\hbar^2$. However the energy shifts associated with the change of
$m$ by 1 are of the order $\hbar$; this discrepancy is due to
singularity of the centrifugal term at $\rho=0$. The appropriate
Gutzwiller expansions in this case use the set of classical
periodic orbits calculated for $L_z=0$, i.e., $m$-independent. The
$m$-dependence of the spectra is taken into account via an
appropriate phase added to the Maslov phase; the phase increment
occurs each time  the orbit hits the axis $\rho=0$, see the
reviews \cite{Fried89, Haseg89}. Since this additional phase is
pseudo-random, the cross-correlation functions between the spectra
with different $m\sim 1$ vanish due to the same mechanism as
between the spectra of different parity in the discrete case.

Let us now turn to the case of $|m|\gg 1$; more precisely consider
the semiclassical limit $\hbar \rightarrow 0,m\rightarrow \infty $
such that the angular momentum projection $L_{z}=\hbar m$ remains
constant. In this case the centrifugal barrier $W$ would not let
the particle approach the $z$ -axis; the singularity at $\rho =0$
will thus be deep in the classical shadow. The change of  $m$ by
$1$ brings about change of the Hamiltonian of the order $\hbar$,
namely by $\Delta H_{m}=\Delta W=\hbar \partial W/\partial L_{z}$.
Before comparing the eigenvalues at different $m$ one must
subtract the overall drift of the spectrum upwards caused by the
increase of the centrifugal energy $W$.  Note that this issue is
similar to the one caused by the short periodic orbits in the
discrete case: it arises from a difference in the one-particle
densities of the two systems, and is an issue which can also be
settled by means of an appropriate unfolding procedure. This
quantity can be evaluated semiclassically as the derivative of its
average over the energy shell,
\begin{equation*}
\left\langle W\right\rangle =\frac{\int d\Gamma \delta \left(
E-H_{m}\right)
W}{\int d\Gamma \delta \left( E-H_{m}\right) }
\end{equation*}%
where $d\Gamma $ is the element of the phase space. Denoting by
$S_{0}$ a typical classical action within  the given (highly excited) energy range,
we can introduce a dimensionless small parameter $\xi =\hbar
/S_{0}$ and write
\begin{eqnarray*}
H_{m+1} &=&H_{m}+\hbar \left( \frac{\partial W}{\partial
L_{z}}-\left\langle
\frac{\partial W}{\partial L_{z}}\right\rangle \right) \equiv
H_{m}+\xi
H^{\prime }, \\
H^{\prime } &=&S_{0}\left( \frac{\partial W}{\partial
L_{z}}-\left\langle \frac{\partial W}{\partial L_{z}}\right\rangle
\right) \,.
\end{eqnarray*}

To appreciate the impact of the perturbation associated with $%
H_{m}\rightarrow H_{m+1}$ consider the general problem of
parametric correlation in the family of Hamiltonians $H_{\xi
}=H_{0}+\xi H^{\prime }$ where $\xi $ is small and dimensionless;
the Hamiltonians $H_{0},H^{\prime }$ are assumed to have finite
classical counterparts. Let us consider the cross form
factor between the spectra of $H_{m}$ and $H_{m+1}$ in the
semiclassical domain. We shall assume that $\xi $ is so small that
the periodic orbits of $H_{m},H_{m+1}$ practically coincide; only
their action difference has to be taken into account since it is
referred to $\hbar $.

The additional classical action due to the perturbation $H^{\prime
}$ over a
 stretch of the orbit with duration from $t_{1}$ to $t_{2}$ would be%
\begin{equation*}
\Delta S\sim \xi \int_{t_{1}}^{t_{2}}H^{\prime }dt.
\end{equation*}%
Now the well-established approach is to model the accumulation of
$\Delta S$ on a trajectory of a chaotic system by a Gaussian
stochastic process \cite{Richt00}; below
 $\left\langle \left\langle \ldots \right\rangle \right\rangle$
denotes averaging over such a process . This can be understood if
we mentally divide the periodic orbit into pieces of duration
larger than the Lyapunov time $T_{L}=\lambda ^{-1}$ where $\lambda
$ is the Lyapunov constant; the action differences accumulated at
two such stretches would be uncorrelated. \ Let $t=t_{2}-t_{1}\gg
T_{L}$ and assume that there is no systematic action increment,
$\left\langle\left\langle \Delta S\right\rangle\right\rangle =0$.
The averaged second moment of $\Delta S$ will grow proportional to
$t$ ,
\begin{equation*}
\left\langle\left\langle \Delta S^{2}\right\rangle\right\rangle =a\xi ^{2}t
\end{equation*}%
where $a$ is a purely classical, system-specific, parameter; its
dimensionality is squared action over time. Remembering that in  a Gaussian
process $\left\langle \left\langle \exp \left( i\Phi \right)
\right\rangle \right\rangle =\exp \left( -\left\langle
\left\langle \Phi ^{2}\right\rangle \right\rangle /2\right) $ we
obtain the average of the exponential of the phase difference
accumulated on a periodic orbit with the period $T_{\gamma },$%
\begin{equation*}
\left\langle \left\langle e^{\frac{i}{\hbar }\Delta S_{\gamma
}}\right\rangle \right\rangle =e^{-a\frac{\xi ^{2}}{2\hbar
^{2}}T_{\gamma }}.
\end{equation*}

Now let us use the semiclassical expression (\ref{semigen}) for
the cross-correlation form factor which we shall now denote
$K_{\xi }\left( \tau \right) $. Limiting ourselves to the diagonal
approximation we would get,
for the time $T=\tau T_{H}$,%
\begin{eqnarray}
K_{\xi ,\mathrm{diag}}\left( \tau \right) &\sim &\frac{1}{\delta
}\sum_{{  T<T_{\gamma }<T+\delta }}%
A_{\gamma
}^{2}\left\langle
\left\langle e^{i\Delta S_{\xi }/\hbar }\right\rangle \right\rangle \\
&=&e^{-a\frac{\xi ^{2}}{2\hbar ^{2}}T}K_{\mathrm{diag}}\left( \tau
\right) .
\end{eqnarray}%
The quantity $K_{\mathrm{diag}}\left( \tau \right) $ is the
autocorrelation form factor in the diagonal approximation, i.e.,
$2\tau $ or $\tau $ depending on the presence or absence of the
time reversal. The perturbation $\xi H^{\prime }$ leads thus to
the decay of the form factor with the characteristic time
$T_{\mathrm{decay}}=2\hbar ^{2}/a\xi ^{2}.$

Applying this result to the correlation between the spectra
belonging to two neighboring values of the quantum number $m$ when
$\xi =\hbar /S_{0}$ we obtain the decay time
\begin{equation*}
T_{\mathrm{decay}}=\frac{2S_{0}^{2}}{a}\,.
\end{equation*}%
This time is classical, i.e., finite in the semiclassical limit,
however it could be much larger than the ballistic time scale,
i.e., the period of the shortest periodic orbits (time on which
correlations are expected anyway).

The corresponding dimensionless $\tau
_{\mathrm{decay}}=T_{\mathrm{decay} }/T_{H}$ turns into zero in in
the limit $\hbar\to 0$.  We therefore find that the
cross-correlation between spectra corresponding to different
symmetries vanishes in the universal regime. Considering that the
periodic orbit expansions of the type (\ref{semigen})--- and above
all the averaging procedure over large numbers of periodic orbits
having nearly the same action---are justified only for times
exceeding the Ehrenfest time $T_{E}\sim \lambda^{-1} \ln
S_{0}/\hbar$, the vanishing of the cross-correlation only holds
for times larger than $T_{E}$. For times shorter than $T_{E}$, and
hence in particular for times of the order of classical times,
correlations are expected to arise, based on the fact that both
symmetry sectors have the same periodic orbits. Such correlations
will depend on the specific features of the periodic orbits and
will thus be highly system-dependent. A possible connection with a
recent discussion of the transition between short time effects and
the universal regime \cite{Smith10} is left for future
investigation.

\section{Conclusions}
In this paper we have given a justification to the common
assumption that spectra corresponding to different irreducible
representations of a symmetry group of a given quantum chaotic
problem are independent in the semi-classical limit. More
precisely we have shown in the semi-classical limit that the
two-point cross correlation vanishes, a property also sometimes
called weak independence. For the specific case of a mirror
symmetry we showed that the cross correlation function is zero not
only in the diagonal approximation and for short times. The non
diagonal terms of the trace formula were also evaluated and the
semi-classical form of the spectral determinant was used to show
that our result holds also beyond the Heisenberg time. For
arbitrary discrete symmetries and for continuous symmetries we
limited our argument to the diagonal approximation, but the
arguments given for mirror symmetries carry over.

An potentially interesting line for future work appears in
continuous symmetries, where the trace formulae reveal an unusual
correlation on a classical time scale. The approximation used is
clearly not valid on this time scale, but the result might still
be a hint, that such correlations exist and are different from the
obvious ones resulting from short orbits.

\section*{Acknowledgements}
 We gratefully acknowledge helpful
discussions with T. Guhr, F. Haake, C. Jung, L. Kaplan, T. Prosen
and R. Schutzhold. Funding from the DFG Sonderforschungsbereich TR
12 as well as under projects IN 114310 from PAPIIT, DGAPA UNAM and
\#79613 of CONACyT is gratefully acknowledged.

\section*{References}
\bibliographystyle{unsrt}
\bibliography{pbraun2}
\end{document}